# CI-Design #11: Hijacking

*Hijacking online reviews: sparse manipulation and behavioral buffering in popularity-biased rating systems*

*Itsuki Fujisaki[1], Kunhao Yang[2]
[1]College of Knowledge and Library Science, School of Informatics, Tsukuba University, Ibaraki, Japan
[2]College of Engineering, Shibaura Institute of Technology; Tokyo, Japan
*Corresponding authors: Itsuki Fujisaki
Email: bpmx3ngj@gmail.com +81-80-3022-6288 (I.F)

**Abstract**

Online reviews and recommendation systems help users navigate overwhelming choice, but they are vulnerable to self-reinforcing distortions. This paper examines how a single malicious reviewer can exploit popularity-biased rating dynamics and whether behavioral heterogeneity in user responses can reduce the damage. We develop a minimal agent-based model in which users choose what to rate partly on the basis of currently displayed averages. We compare broad attacks that perturb many items with sparse attacks that selectively boost low-quality items and suppress high-quality items. Additional analyses not shown here indicate that sparse attacks are substantially more harmful than broad attacks because they better exploit popularity-based exposure. The main text then focuses on sparse attacks and asks how their effects change as the fraction of contrarian users increases. Three results stand out. First, attack-induced damage is strongest when prior honest reviews are scarce, revealing a transition from a fragile low-information regime to a more robust high-information regime. Second, sparse attacks are especially effective at artificially promoting low-quality items. Third, moderate contrarian diversity partially buffers these distortions, primarily by suppressing the rise of low-quality items rather than fully restoring high-quality items to the top. The findings suggest that recommendation robustness depends not only on attack detection and predictive accuracy, but also on review density, popularity feedback, and user response heterogeneity.

**Keywords**

Online reviews; Recommendation systems; Social influence; Shilling attacks; Contrarian behavior; Agent-based simulation

## Introduction

Recommendation systems have become a basic part of everyday information processing. People rely on them when choosing a hotel, a restaurant, a book, a video, a news article, or a piece of software. In each of these settings, the same underlying problem appears in slightly different form. The number of available options is too large for users to inspect one by one, and the cost of gathering reliable information independently is often too high. Ratings, rankings, review counts, and recommendation lists therefore function as practical devices for reducing uncertainty and narrowing the search space. They help users decide what deserves attention and what can safely be ignored. In this sense, recommendation systems are not merely computational conveniences. They are information environments that distribute visibility, organize collective traces, and shape the conditions under which judgments are formed. This is why they have been studied so extensively in information science, recommender-systems research, and computational social science [1, 2]. Their purpose is not only to predict what a user will like, but also to make collective information usable at scale.

Online review systems occupy a particularly important place within this landscape. Compared with recommendation lists that are generated entirely inside a model, reviews and ratings also serve as public records of evaluation. They affect trust, discovery, and perceived quality at the same time. Research on online reputation systems has long emphasized their practical value for reducing uncertainty in environments where direct inspection of quality is costly or impossible [3, 4]. Reviews can influence real economic outcomes as well. They affect product visibility, consumer choice, and even firm revenues [5, 6]. For precisely this reason, they have become central to many digital platforms. A review system is not just a place where users leave opinions after the fact. It is often part of the platform's primary interface for guiding later users toward some options and away from others.

Yet the very same properties that make these systems useful also make them fragile. Most recommendation environments depend, directly or indirectly, on the traces left by previous users. Prior clicks, downloads, ratings, purchases, and reviews are not passive records. They are transformed into signals that affect what later users see. Once this happens, the system no longer simply aggregates independent judgments. It begins to feed past outcomes back into future exposure. A substantial body of work has shown that this feedback can be socially consequential. In their well-known artificial cultural market experiments, Salganik, Dodds, and Watts showed that when participants could see what others had chosen, inequality and unpredictability increased sharply [7]. Small early differences became magnified, and market outcomes became less tightly linked to intrinsic appeal. Salganik and Watts later showed that even minor early perturbations can become self-fulfilling [8]. These studies matter because they reveal a general problem. Recommendation-like feedback does not merely report collective preference. It can reshape collective preference by directing later attention toward what already appears successful.

Subsequent work has shown that these distortions may arise even when the initial intervention is extremely small. Muchnik, Aral, and Taylor found that tiny manipulations of early social ratings can have large downstream consequences [9]. A small artificial boost at the beginning can trigger a cascade of later positive judgments. The important point is not simply that people imitate others. It is that early signals enter later visibility, and later visibility enters later judgment. Once that feedback loop is in place, recommendation systems stop being neutral aggregators of information and become active mediators of informational opportunity. This is especially important in digital environments, where a user's chance to evaluate an option often depends on whether a platform exposes that option at all. Recommendation systems therefore sit at the core of a broader information-processing problem. They do not merely help users consume information more efficiently. They also determine which information is encountered, which information is ignored, and which information becomes credible because it has been made socially legible.

This issue has been studied more directly in recent work on popularity bias. Ciampaglia, Nematzadeh, Menczer, and Flammini showed that algorithmic popularity bias can either hinder or promote quality, depending on the surrounding structure of search and attention [10]. This is an important correction to the overly simple claim that popularity is always harmful. Popularity cues may, under some conditions, help users coordinate on good content. Under other conditions they may suppress exploration, amplify noise, and distort quality signals. Similar concerns appear in work on recommender-system bias. Fleder and Hosanagar showed that collaborative filtering systems can promote concentration and create artificial blockbusters [11]. Chaney, Stewart, and Engelhardt demonstrated that recommendation algorithms can create self-reinforcing loops between user behavior and training data, thereby producing algorithmic confounding [12]. More recent studies have connected these processes to popularity bias, exposure inequality, and mainstream-taste bias in recommender systems [13–15]. The key lesson is that the effects of popularity are regime-dependent. A recommendation rule that works reasonably well when many reliable evaluations already exist may work badly when information is sparse or when early signals are noisy. The design problem is therefore not only one of predictive accuracy. It is also one of informational integrity. A recommendation environment should not merely guess what a user is likely to click. It should preserve the conditions under which meaningful signals can emerge and remain visible.

One reason this problem is so challenging is that recommendation systems may generate distortions even in the absence of any explicitly malicious actor. A growing literature shows that review and ranking systems are vulnerable to endogenous crowd effects. Le Mens, Kovács, Avrahami, and Kareev demonstrated that online ratings can be undermined by endogenous crowd formation [16]. If items with favorable displayed averages attract more evaluations while items with unfavorable averages attract fewer, then the crowd that accumulates around each item is not random. It is selected by the platform's own display mechanism. Denrell and Le Mens developed a related argument in their work on information sampling and collective illusions [17]. Aggregate distortions do not require irrational evaluators or biased individual judgments. They may arise because some alternatives are repeatedly sampled while others remain largely unseen. Germano, Gómez, and Le Mens extended this logic to ranked settings and showed that popularity-based interfaces can produce "few-get-richer" dynamics, in which attention concentrates disproportionately on a small subset of items [18]. Analytis, Gelastopoulos, and Stojic further formalized ranking-based rich-get-richer processes and clarified how rank reinforcement can reshape exposure over time [19]. Taken together, this body of work suggests that recommendation systems can be informationally fragile from the inside. Even before any attacker appears, the coupling of ranking, visibility, and evaluation may generate systematic distortion.

If recommendation systems are already vulnerable under benign conditions, then intentional manipulation becomes even more serious. This problem has long been studied in research on fake reviews and shilling attacks. Lappas, Sabnis, and Valkanas showed that fake reviews can influence online visibility, meaning that manipulation can affect not only local ratings but also the broader exposure structure of a platform [20]. In recommender-systems research, similar concerns are typically framed as profile-injection or shilling attacks, in which malicious actors submit strategically designed evaluations to push some items upward or nuke others downward [21]. More recent work shows that review-based recommendation models may be especially vulnerable because reviews do not just produce visible scores. They also shape internal user and item representations, which makes sparse but carefully designed manipulation surprisingly consequential [22]. These findings are central for information science because they show that the trustworthiness of online information systems depends not only on whether users tell the truth, but also on whether the system can resist having small distortions amplified into large informational consequences.

A particularly important contribution in this direction is the work of Pescetelli, Barkoczi, and Cebrian, who showed that even a single malicious bot can influence collective outcomes indirectly through a recommender system [23]. Their result is striking because it shifts the problem away from sheer scale. It is easy to imagine that many coordinated bots can distort an online environment. It is more revealing to show that a single strategically placed malicious actor may matter if the platform amplifies its influence. In a different domain, Berdoz, Rugli, and Wattenhofer make a related point by showing that even one malicious participant can sharply reduce the success of a cooperative system [24]. These studies suggest that "one bad actor" is not merely a metaphor. Under the right conditions, it can be an empirically meaningful description of systemic fragility. What remains underexplored, however, is how this logic plays out in popularity-biased review systems, where visibility and evaluation are tightly coupled and manipulation may be amplified by the same dynamics that make recommendation effective in the first place.

The present study addresses that question in a minimal model of online review dynamics. We ask how a single malicious reviewer can distort a popularity-biased rating system and under what conditions that distortion is amplified or absorbed. We compare two types of attack. A broad attack perturbs many items at once, rating good items downward and bad items upward across the board. A sparse attack instead targets only a small subset of items, selectively boosting a few low-quality options and suppressing a few high-quality ones. This distinction turns out to be crucial. A natural intuition is that broad manipulation should be most destructive because it affects the largest number of items. Our results suggest the opposite. Broad attacks are surprisingly inefficient. Because they touch everything, they partially flatten the very visibility differentials on which popularity amplification depends. Sparse attacks are more dangerous precisely because they are selective. They introduce local distortions that are then magnified by popularity-biased exposure.

This leads to one of the central findings of the paper. The danger posed by a malicious reviewer is strongly dependent on prior review density. When honest reviews are scarce, the system operates in a low-information regime in which early displayed signals carry disproportionate weight. Under those conditions, a sparse intervention can generate large downstream distortions. When honest reviews are already abundant, the same malicious intervention loses much of its force. The system moves into a more robust regime because no single injected evaluation can strongly alter the information environment. This result is theoretically important because it shows that vulnerability is not a fixed property of a platform. It emerges from the interaction between popularity bias and the amount of honest information already accumulated. In other words, the same platform rule can be safe or fragile depending on when in the accumulation process it is applied.

The paper also asks whether recommendation systems contain any endogenous source of protection against such attacks. Here a second literature becomes relevant, one centered on divergence from popularity rather than conformity to it. Behavioral work has repeatedly shown that people do not always follow what is popular. Berger and Heath demonstrated that consumers may actively avoid options that become associated with dissimilar others or outgroups because such options no longer signal a desired identity [25, 26]. Tuk, Verlegh, Smidts, and Wigboldus similarly showed that perceived dissimilarity can produce preference contrast, leading people to move away from others' choices rather than toward them [27]. Classic work on minority influence and nonconformity makes a related point, namely that deviation from majority behavior can preserve informational diversity and resist premature convergence [28, 29]. These studies matter because they establish that contrarian responses are not merely a theoretical convenience. They are empirically observed patterns of behavior.

Recent work by the present authors suggests that such contrarian responses are not merely idiosyncratic noise. In popularity-biased rating systems, moderate mixtures of conformist and contrarian evaluators can improve collective accuracy by redirecting attention toward neglected options [30]. Relatedly, dissimilar sources can sometimes be informationally useful in matters of taste, especially for non-mainstream users [31]. These findings suggest that user heterogeneity in response to social information may play a constructive role in recommendation environments. This paper therefore asks whether the same behavioral diversity that can improve ordinary collective evaluation might also buffer systems against adversarial distortion.

Our results support this possibility, but in a qualified way. Contrarian diversity does not fully restore a manipulated system to an ideal state. Rather, it acts as a partial buffer. Most notably, moderate contrarianism suppresses the artificial promotion of low-quality items more effectively than it restores high-quality items to the top once they have been attacked. This asymmetry is revealing. It suggests that contrarian responses do not simply undo the consequences of manipulation. Instead, they interfere with the pathway through which manipulation becomes socially amplified. In a purely conformist environment, a sparse attack can make a bad item look socially attractive, which then pulls more attention toward it. In a mixed environment, some users systematically resist that popularity cue, limiting the rise of the manipulated item. The protective effect is therefore real, but selective. It is strongest against the social overpromotion of weak items and weaker against the complete restoration of strong items.

This asymmetry also clarifies why building healthy recommendation systems is so difficult. A platform may fail not because its prediction model is inaccurate, but because it creates a visibility regime in which small perturbations are amplified into large distortions. Conversely, a platform may benefit from forms of behavioral diversity that are often treated as inefficiency. The lesson is not that recommendation systems should simply encourage users to behave contrarily. The lesson is that robustness depends on the interaction of ranking rules, prior information density, attack structure, and response heterogeneity. This is fundamentally an information-processing problem. What matters is not only what information exists, but which information becomes visible, how it is sampled, and how people react to the resulting signals. In that sense, the present study is aligned with the concerns of EPJ Data Science. It examines a digital platform process in which local interventions, algorithmic mediation, and collective dynamics jointly determine systemic outcomes.

The paper therefore makes three contributions. First, it extends prior work on bots, fake reviews, and shilling by showing how a single malicious reviewer can exploit popularity-biased rating aggregation without requiring large-scale coordinated manipulation. Second, it identifies a counterintuitive vulnerability: sparse targeted attacks are more dangerous than broad attacks because they better exploit the self-reinforcing dynamics of popularity-based exposure. Third, it shows that behaviorally grounded contrarian responses can partially buffer these attacks, especially by preventing low-quality items from being artificially elevated. More broadly, the paper argues that understanding recommendation robustness requires attention not only to attacks and algorithms, but also to the behavioral diversity of users who respond to popularity cues. Recommendation systems are powerful because they transform collective traces into usable signals. The same transformation is also what makes them vulnerable. Understanding when those systems become fragile, and when behavioral heterogeneity can help stabilize them, is therefore an important task for computational social science and data-driven research on digital platforms.

## Methods

### 2.1 Model overview

We studied the vulnerability of popularity-biased review systems using a minimal sequential simulation model. The basic logic of the model follows recent work on popularity bias, information sampling, and endogenous crowd formation [10, 16–19]. At the same time, it is designed to speak directly to the literature on fake reviews, shilling, and malicious platform manipulation [20–24]. The purpose of the model is not to reproduce the full complexity of a commercial recommender system. Instead, it isolates the smallest ingredients required to ask a focused question: when can one malicious reviewer distort a popularity-biased review system, and under what conditions can behavioral heterogeneity reduce that distortion?

Each simulation run began with K = 50 items. Each item i had an underlying latent quality mu_i, drawn from a common distribution centered on the middle of the rating scale. Honest evaluations were generated as noisy signals of this quality and then clipped to the standard 1–5 star range. This means that honest reviewers were not biased in the content of their ratings. Distortion in the model arose primarily through selective exposure and selective evaluation, rather than through dishonest ordinary users. This assumption follows the logic of endogenous crowd formation models. Even when individual evaluations are unbiased, aggregate ratings can become distorted if the platform changes which items receive attention [16, 17].

The simulation unfolded in two stages. In the first stage, each item received a number of prior honest reviews, denoted m_{text{seed}}. This parameter controlled the thickness of the initial information environment. When m_{text{seed}} = 0, the system began in an extremely sparse-information regime. When m_{text{seed}} was larger, the system started from a more stable base of honest information. In the main figures we focused on four representative levels, m_{text{seed}} in {0, 2, 5, 10}, because these values provide a compact view of the shift from fragile to relatively robust conditions.

In the second stage, additional users arrived sequentially and chose which item to evaluate. Choice probabilities were governed by currently displayed average ratings through a softmax rule. A social influence parameter beta controlled how strongly displayed popularity shaped attention. Higher beta values made users more likely to evaluate items with favorable displayed averages, thereby strengthening conformity to popularity. Lower values weakened this effect. In the main figures we focused on beta in {1.5, 2.5, 4.0}, which span weak, intermediate, and relatively strong popularity feedback.

### 2.2 Honest users and contrarian users

A central concern of the paper was whether behavioral heterogeneity in users' responses to popularity could buffer the effects of manipulation. We therefore distinguished between conformist and contrarian users. Conformist users sampled items in the usual popularity-biased direction: items with higher displayed averages were more likely to receive attention. Contrarian users reversed the sign of this popularity response and were more likely to evaluate items with lower displayed averages. Let q denote the fraction of contrarian users among the sequential arrivals. When q = 0, all users are conformist. As q increases, the system becomes more behaviorally diverse.

This implementation follows the spirit of prior work showing that moderate mixtures of conformist and contrarian behavior can improve collective accuracy in popularity-biased settings [30]. In the present paper, however, the interest is different. We ask whether the same heterogeneity can also make a system more resistant to adversarial manipulation.

### 2.3 Attack structures

The model compared two forms of malicious intervention. In a broad attack, a single malicious reviewer rated many items, pushing low-quality items upward and high-quality items downward across the board. In a sparse attack, the same reviewer submitted only two strategically chosen ratings: a maximal rating for the true lowest-quality item and a minimal rating for the true highest-quality item. The broad attack corresponds to an intuitive but diffuse form of sabotage. The sparse attack corresponds to a selective manipulation that attempts to exploit the system's own self-reinforcing dynamics.

Preliminary analyses comparing these two attack families showed that broad attacks were consistently less harmful than sparse attacks across the parameter regimes considered. Because the targeted attack is therefore the more consequential and theoretically interesting case, the main text focuses on sparse attacks. Comparisons between broad and sparse attacks, together with heat-map summaries over a wider parameter grid, are currently being consolidated as supplementary analyses. The qualitative result is clear: broad attack is surprisingly inefficient, whereas sparse attack is much more capable of hijacking popularity-based review dynamics.

### 2.4 Outcome variables

We tracked three complementary outcome variables. The first was the attack-induced change in aggregate error, Delta text{RMSE}, defined as the difference between the root mean squared error under attack and the corresponding error without attack under the same parameter settings. This is the broadest measure of system-level informational damage.

The second outcome was best-item demotion. This measured how far the true highest-quality item fell in the final ranking under attack. It captures the possibility that manipulation suppresses the visibility of genuinely strong items.

The third outcome was worst-item promotion. This measured how far the true lowest-quality item rose in the final ranking under attack. It captures a complementary and especially consequential possibility: that sparse manipulation does not simply hide good options, but actively creates the appearance that poor options are socially validated.

The three outcomes were chosen because they speak to different aspects of robustness. Aggregate error indicates whether the system's overall informational quality deteriorates. Best-item demotion indicates whether valuable information is lost from the top of the ranking. Worst-item promotion indicates whether weak information is made artificially prominent.

### 2.5 Parameter settings and simulation runs

The main text reports simulations over the four levels of prior honest reviews (m_{text{seed}} = 0, 2, 5, 10), three levels of popularity feedback (beta = 1.5, 2.5, 4.0), and contrarian fractions ranging from $q = 0$ to $q = 0.5$. A fixed number of additional honest evaluations was then generated for each item. For every condition, outcomes were estimated by repeated Monte Carlo simulations and summarized by their mean values. Additional comparisons over broader parameter grids yielded similar patterns and are being organized for supplementary presentation.

## Results

### 3.1 Sparse attack increases aggregate error, especially when prior reviews are scarce

Figure 1 shows the broadest result of the paper. Sparse attack substantially increases aggregate error when prior honest reviews are scarce, but this vulnerability rapidly declines as the system becomes better anchored by honest information. When m_{text{seed}} = 0, the attack produces the largest increase in RMSE across all three values of beta. This is the most fragile regime. Here early displayed averages are weakly constrained, and later users rely heavily on those early signals when choosing what to evaluate. Under such conditions, a single malicious intervention is sufficient to distort the information environment in a visible way.

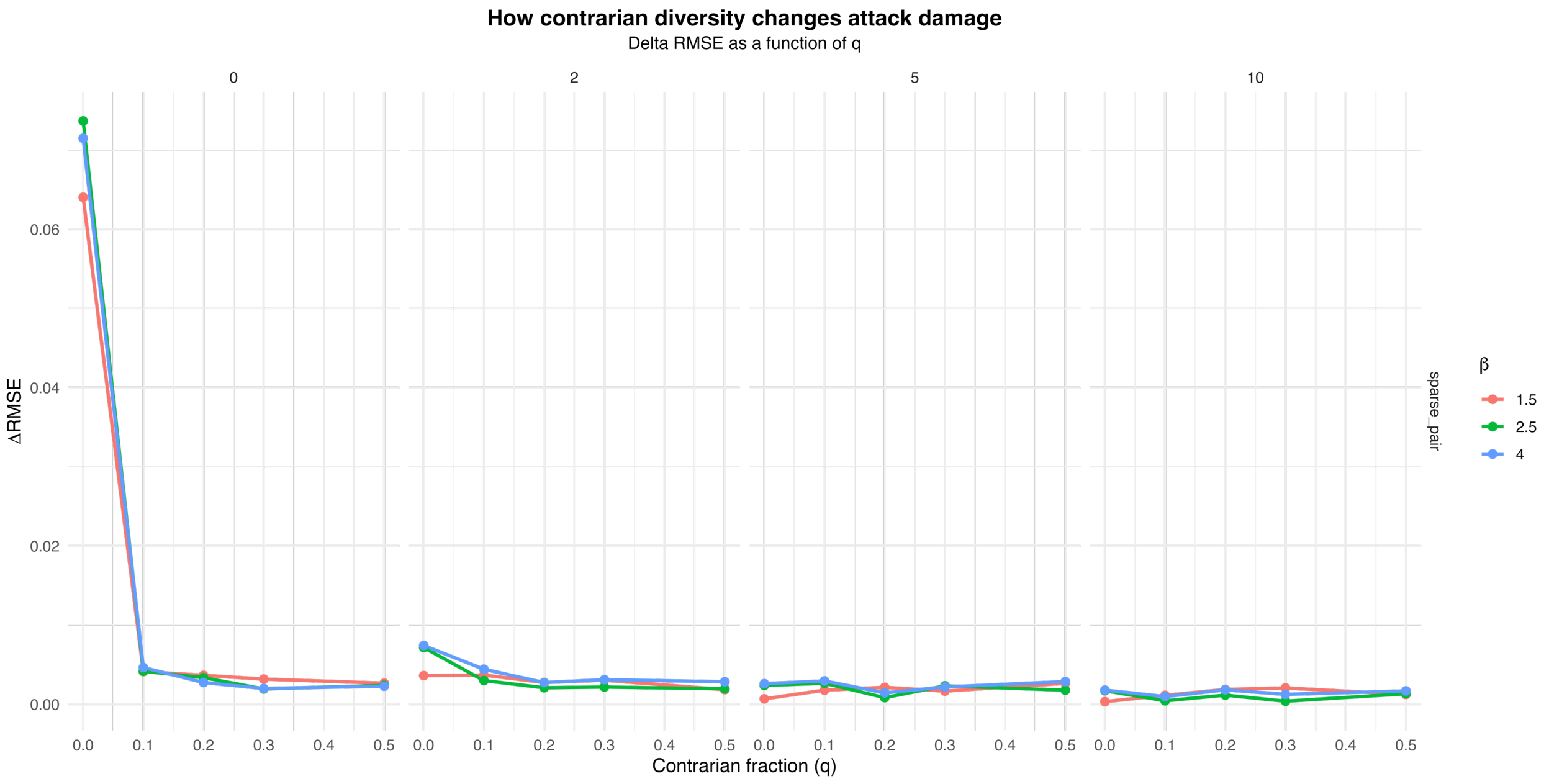

**Figure 1. Attack-induced change in collective error as a function of contrarian fraction.Mean DeltaRMSE under sparse attack, plotted against the fraction of contrarian users q. Panels correspond to the number of prior honest seed reviews per item (m_{mathrm{seed}} = 0, 2, 5, 10), and colors indicate the strength of popularity-biased exposure (beta = 1.5, 2.5, 4.0). Attack-induced error is largest when prior honest information is scarce and declines rapidly as review density increases. Increasing contrarian diversity reduces the damage, with most of the buffering occurring between q = 0 and q = 0.1.**

As prior honest reviews accumulate, however, the same attack loses force. At m_{text{seed}} = 2, the attack still raises RMSE, but the effect is much smaller. By m_{text{seed}} = 5 and especially m_{text{seed}} = 10, the additional damage to RMSE is close to negligible. The implication is straightforward. A review system's vulnerability is not fixed. It depends strongly on its review density. Systems with very thin review histories are highly sensitive to targeted manipulation. Systems with even modest prior accumulation become substantially more robust.

Figure 1 also shows that contrarian diversity partially buffers this damage. The strongest reduction in Delta text{RMSE} occurs when q moves from 0 to 0.1. Larger increases in q continue to reduce the damage, but much more gradually. By q = 0.3 or 0.5, the curves have largely flattened. This means that the buffering effect of contrarian behavior is real, but the system does not require a large contrarian majority for most of the benefit to appear. Even a modest amount of divergence from popularity already reduces the ability of sparse attack to propagate through the system.

### 3.2 Contrarian diversity strongly suppresses the promotion of low-quality items

Figure 2 reveals where this buffering effect is most pronounced. Sparse attack is especially effective at promoting the true lowest-quality item, and contrarian diversity substantially reduces this promotion. When m_{text{seed}} = 0, the worst item can move upward by a large number of ranks when all users are conformist. This artificial rise becomes noticeably smaller as q increases. A similar, though weaker, pattern is visible when m_{text{seed}} = 2. Once m_{text{seed}} reaches 5 or 10, worst-item promotion becomes modest overall and the scope for further buffering correspondingly shrinks.

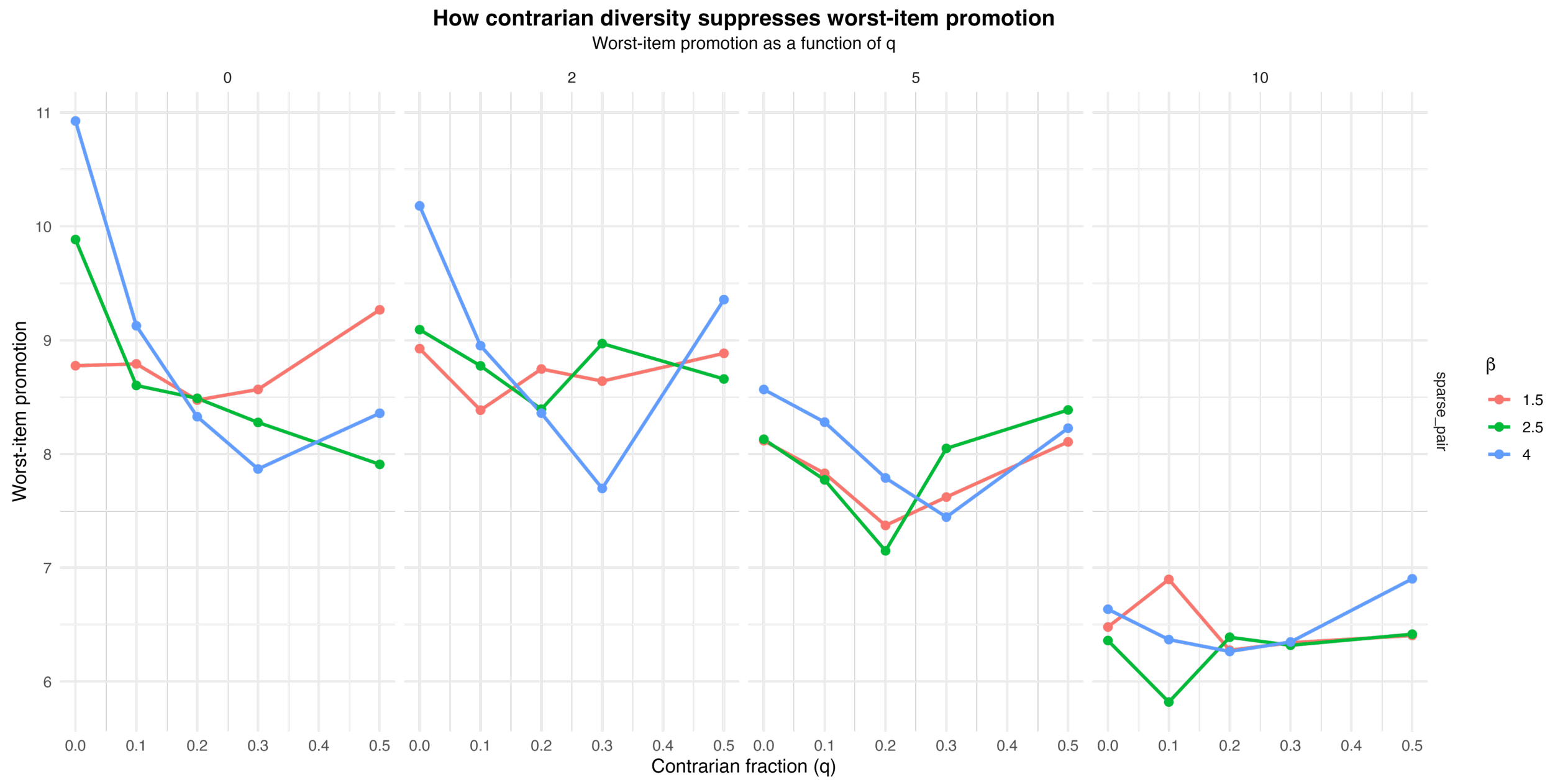

**Figure 2. Contrarian diversity suppresses the promotion of low-quality items under sparse attack. Mean upward rank shift of the true lowest-quality item under sparse attack, plotted against the fraction of contrarian users q. Panels correspond to prior honest seed reviews per item, and colors indicate the strength of popularity-biased exposure. Sparse manipulation is particularly effective at artificially promoting low-quality items when prior reviews are scarce. This promotion is substantially reduced as contrarian diversity increases, especially in the lowest-information regime.**

This figure is especially important because it identifies the clearest mechanism behind the aggregate-error result. Contrarian diversity does not simply make the system “better” in an abstract sense. It specifically reduces the chance that a low-quality item will become socially amplified after being strategically manipulated. In a purely conformist environment, a sparse attack can make a weak item look unusually attractive, and later users reinforce that impression by continuing to evaluate it. In a mixed environment, some users systematically resist that local popularity cue, limiting the upward drift of the manipulated item.

### 3.3 Protection against demotion of the best item is weaker and less regular

Figure 3 shows a more complicated pattern. Sparse attack also pushes the true best item downward, but the buffering effect of contrarian diversity is weaker and less regular than in the case of worst-item promotion. When m_{text{seed}} = 0, the best item is clearly demoted under attack, and increasing q reduces this effect to some degree. But once m_{text{seed}} reaches 2 or more, the curves are flatter and more uneven. The system becomes less vulnerable overall, but the role of contrarianism is harder to characterize by a simple monotonic relation.

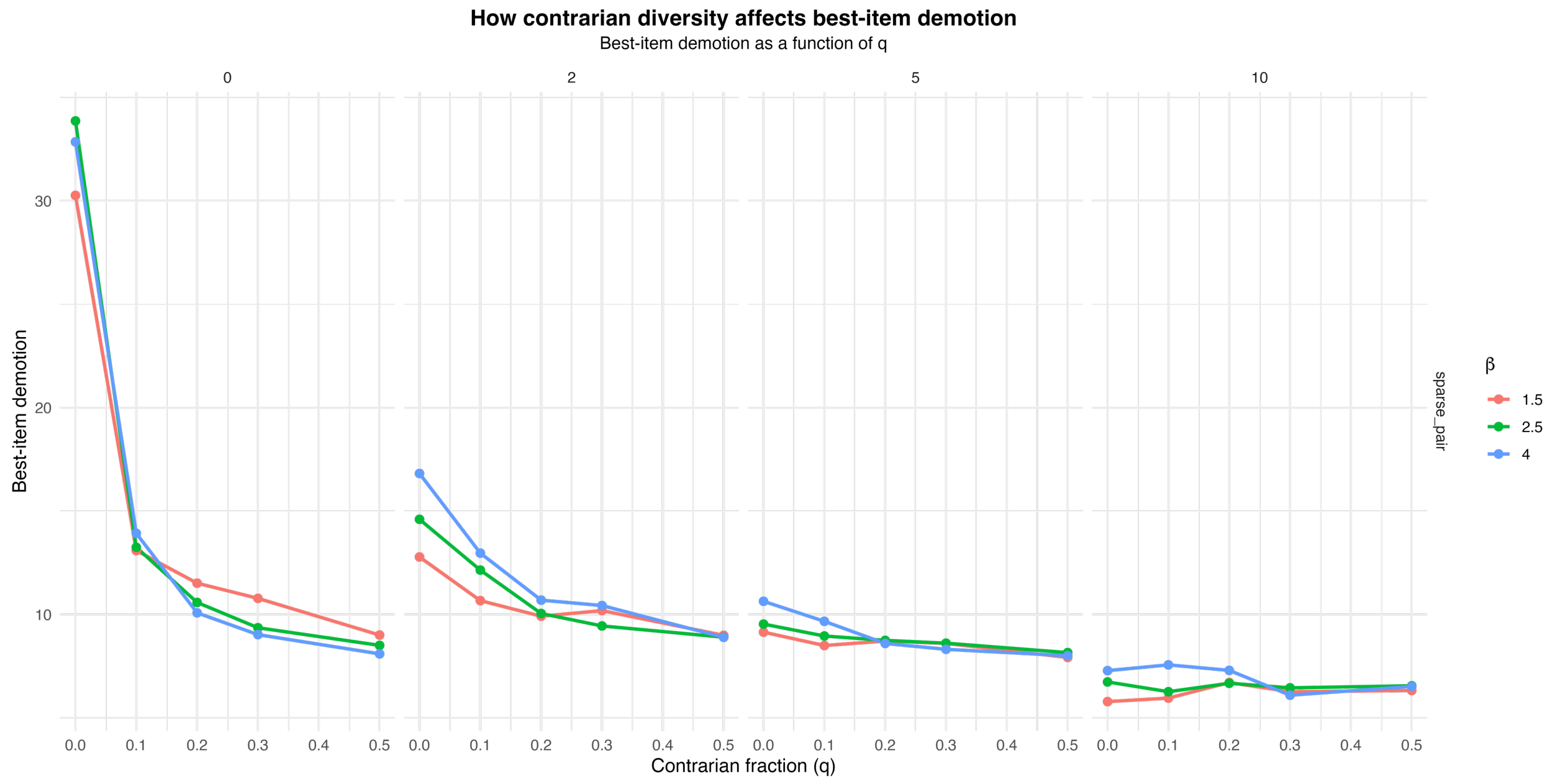

**Figure 3. Contrarian diversity provides weaker protection against the demotion of high-quality items. Mean downward rank shift of the true highest-quality item under sparse attack, plotted against the fraction of contrarian users q. Panels correspond to prior honest seed reviews per item, and colors indicate the strength of popularity-biased exposure. Sparse attack also demotes high-quality items, but the buffering effect of contrarian diversity is weaker and less regular than in the case of low-quality item promotion. This asymmetry suggests that contrarian users more effectively prevent the artificial rise of poor items than fully restore the visibility of strong ones.**

This asymmetry is theoretically informative. It suggests that contrarian diversity is more effective at preventing the artificial rise of poor items than at restoring strong items once they have already been pushed downward. A likely reason is that contrarian behavior changes the direction of attention but does not necessarily "rediscover" an item once it has become less visible. In other words, contrarians can resist a locally inflated popularity signal more easily than they can reconstruct a suppressed high-quality signal.

### 3.4 Additional analyses

Additional analyses currently being finalized for the supplementary material strengthen this interpretation in three ways. First, direct comparisons between broad and sparse attacks indicate that broad attacks are consistently less harmful than sparse attacks. Second, regime maps over the full m_{text{seed}} times beta surface confirm the same fragile-to-robust transition suggested by the main figures. Third, baseline analyses without malicious intervention reproduce the earlier finding that moderate mixtures of conformist and contrarian behavior can improve collective accuracy in popularity-biased rating systems [30]. Together, these extensions reinforce the conclusion that contrarian diversity functions not only as a source of ordinary collective accuracy, but also as a partial buffer against adversarial distortion.

## Discussion

The main contribution of this paper is to show that vulnerability in online review systems is not simply a matter of whether manipulation exists, but of how manipulation interacts with the system's own visibility dynamics. A single malicious reviewer does not distort a platform equally under all conditions. The effect depends on the structure of the attack, the amount of honest information already present, and the behavioral composition of later users. This is an important shift in perspective. Much of the literature on fake reviews, shilling, and bots treats manipulation primarily as an adversarial input problem [20–24]. Our results suggest that this is only part of the story. What matters at least as much is the information regime into which the manipulation is inserted. The same malicious act can be absorbed almost completely when review density is high, yet become highly consequential when review density is low and popularity-biased exposure is already active.

This perspective helps explain why sparse attacks outperform broad attacks in our simulations. At first glance, a broad attack seems as though it should be more destructive because it distorts more of the system. Yet its very breadth turns out to be a weakness. Because it perturbs many items at once, it partly flattens the relative differences on which popularity amplification depends. Sparse attack does the opposite. It leaves most of the system untouched while creating a few locally strong distortions. These local distortions are then magnified by the same feedback process that normally helps popular items accumulate attention. The attack succeeds not by overwhelming the system, but by exploiting its own endogenous logic. This result complements prior work showing that fake reviews can distort visibility [20] and that even one malicious bot can alter collective outcomes through a recommender system [23]. The present paper adds that selectivity is a crucial ingredient of effective manipulation in popularity-biased review systems.

A second key result is that prior review density creates a practically meaningful distinction between fragile and relatively robust regimes. When the system is thinly informed, early displayed averages carry disproportionate weight. Under those conditions, even a single malicious review can produce a downstream effect large enough to change later exposure and later evaluation. Once a modest stock of honest reviews exists, however, the same attack becomes much less influential. This finding is consistent with earlier work on social influence and endogenous crowd formation, where small initial signals matter most when a system has not yet accumulated enough information to stabilize itself [7–10, 16–19]. It also reinforces the broader lesson that recommendation systems should not be evaluated only in a steady-state sense. Their vulnerability may be concentrated in the early stages of information accumulation, when local perturbations are most likely to be amplified.

The distinction between aggregate error and rank distortion is also important. Figure 1 shows the system-level effect on RMSE, but Figures 2 and 3 show that the content of that distortion matters. The most robust effect concerns the promotion of low-quality items. Sparse attack is particularly effective at giving weak items a misleading appearance of social legitimacy, and contrarian diversity is particularly effective at reducing this effect. This matters because online review systems are not just ranking abstract vectors. They are structuring real visibility. A weak item that rises in the ranking becomes easier to encounter, easier to trust, and more likely to accumulate further evaluations. In information-management terms, attack is therefore not only an accuracy problem. It is also a visibility problem. The system fails not just because it estimates quality badly, but because it redistributes attention in ways that make poor information seem socially validated.

The asymmetry between worst-item promotion and best-item demotion is especially revealing. Our results show that sparse attack more reliably inflates bad items than it suppresses good ones, and that contrarian diversity more reliably blocks the former than corrects the latter. This is not a nuisance result. It suggests that contrarianism acts more as a brake on runaway amplification than as a perfect repair mechanism. Once a high-quality item has already been pushed downward and deprived of visibility, a mixed user population does not automatically send attention back toward it. By contrast, if a low-quality item has been made to look socially attractive, contrarian users are structurally well positioned to resist that attraction. In that sense, behavioral diversity changes the pathway through which manipulation becomes amplified rather than symmetrically undoing all of its consequences.

This interpretation connects directly to the behavioral literature on divergence from popularity. Berger and Heath [25, 26] showed that consumers may actively avoid options associated with dissimilar others, and Tuk et al. [27] showed that perceived dissimilarity can generate preference contrast rather than assimilation. These findings matter here because they allow contrarianism to be interpreted as a plausible behavioral tendency rather than a purely formal device. Recent simulation work by Fujisaki and Yang [30, 31] suggested that such divergence may improve collective accuracy under benign conditions. The present study extends that argument into an adversarial setting. Contrarian diversity does not merely enrich the information environment in ordinary circumstances. It also increases resistance to certain forms of manipulation. At the same time, our results are more nuanced than a simple claim that "contrarians are good." The benefit is strongest against the artificial popularization of low-quality items and weaker against the recovery of high-quality items that have already lost visibility. That limited, asymmetric effect makes the broader claim more credible rather than less.

These results also clarify the relationship between this study and recent work on bots and recommendation systems. Pescetelli et al. [23] showed that a single bot can shape collective outcomes indirectly through algorithmic mediation. Our results support the same general intuition, but in a more specific environment. We show not only that one bad actor can matter, but also how and when it matters in popularity-biased review systems. The answer is that sparse attack matters most when inserted into a low-information regime where popularity-based exposure can magnify local perturbations. This interactional view is important. It suggests that the relevant unit of analysis is not simply the attacker, but the combination of attacker, ranking logic, and user response heterogeneity. That may help explain why some review systems appear surprisingly robust in practice while others seem vulnerable to relatively small distortions.

From a design perspective, the implications are immediate. A platform that foregrounds popularity cues when honest review density is still very low may be unintentionally creating the conditions under which sparse manipulation is most effective. One possible remedy is to delay or downweight popularity signals until a minimum stock of honest information is present. Another is to diversify exposure more aggressively in early phases so that local perturbations are less likely to become self-reinforcing. A third is to focus monitoring not only on broad suspicious patterns, but also on sparse, strategically placed distortions, since our results suggest that these may be more damaging than broader attacks. More broadly, the present findings support the view that recommendation robustness should not be understood solely as a problem of model accuracy or attack detection. It is equally a problem of visibility management, exposure design, and response heterogeneity.

Of course, the model is intentionally minimal, and that minimality sets limits on what can be claimed. We examined a single malicious reviewer rather than coordinated adversarial populations. We focused on seed-stage intervention rather than later attacks or repeated manipulation. We treated contrarianism as a stable response tendency rather than a context-sensitive or strategic one. We also worked in a stylized environment with known latent quality, which is necessary for clear evaluation but not directly observable in real platforms. These simplifications were deliberate, because the purpose of the paper was to isolate a mechanism rather than reproduce every aspect of a commercial recommender system. Even so, they point to clear directions for future work. Ongoing supplementary analyses already indicate that broad attacks are less harmful than sparse attacks and that regime maps over the full parameter surface yield the same qualitative picture as the main figures. Further extensions could examine attack timing, multiple attackers, richer rating-generation processes, or semi-synthetic validation using real review trajectories.

What the present study establishes, however, is already important. Online review systems are vulnerable not simply because they can be attacked, but because their own popularity dynamics can amplify carefully placed local distortions. Sparse attacks matter more than broad attacks because they exploit, rather than overwhelm, the system's self-reinforcing logic. This vulnerability is strongest when honest information is thin and much weaker when review density is already moderate. Contrarian diversity does not solve the problem completely, but it does make the system meaningfully harder to hijack, especially by reducing the artificial rise of poor-quality items. In this sense, the paper argues for a broader understanding of recommendation robustness. Robustness is not only about filtering out malicious input after the fact. It is also about designing information environments in which small distortions cannot so easily become large social facts.

## Data availability

Code and processed outputs will be made available in OSF.

## Acknowledgments

None.

## Declaration of interests

The authors declare no competing interests.

## Funding statement

This research received no external funding.

## Ethical approval and informed consent statements

This study did not involve human participants or animals. Ethical approval and informed consent were therefore not required.